\documentclass{sig-alternate-05-2015}

\usepackage{algorithm}
\usepackage{algpseudocode}
\usepackage{makeidx}  
\usepackage{booktabs}
\usepackage{url}
\usepackage{amssymb}
\usepackage{amsmath}
\usepackage{color}
\usepackage{epstopdf}
\usepackage{balance}

\newcommand{\BigO}[1]{\ensuremath{\mathcal{O}(#1)}}

\begin{document}

\setcopyright{acmcopyright}
\conferenceinfo{SoICT 2015,}{December 03-04, 2015, Hue City, Viet
Nam}
\isbn{978-1-4503-3843-1/15/12}\acmPrice{\$15.00}
\doi{http://dx.doi.org/10.1145/2833258.2833302}

%

\title{Minimizing Total Busy Time for Energy-Aware Virtual Machine Allocation Problems}

\numberofauthors{2}
\author{
\alignauthor Nguyen Quang-Hung\\
		\affaddr{Faculty of Computer Science and Engineering} \\
		\affaddr{HCMC University of Technology} \\
		\affaddr{Ho Chi Minh City, Vietnam} \\
		\email{hungnq2@cse.hcmut.edu.vn}
\alignauthor Nam Thoai\\
		\affaddr{Faculty of Computer Science and Engineering} \\
		\affaddr{HCMC University of Technology} \\
		\affaddr{Ho Chi Minh City, Vietnam} \\
		\email{nam@cse.hcmut.edu.vn}
}

\maketitle              

\begin{abstract}
This paper investigates the energy-aware 
virtual machine (VM) allocation problems in clouds along characteristics:
multiple resources, 
fixed interval time and non-preemption of virtual machines.
Many previous works have been proposed to use a minimum number of physical machines;
however, this
is not necessarily a good solution to minimize total energy consumption in the VM placement with multiple resources, fixed interval time and non-preemption.
 We observed that minimizing the sum of total busy time of all
physical machines implies minimizing total energy consumption of physical machines.
In addition to, if mapping of a VM onto physical machines have the same total busy time then 
the best mapping has physical machine's remaining available resource minimizing.
Based on these observations,
we proposed heuristic-based EM algorithm to solve the energy-aware VM allocation with fixed starting time and duration time.
 In addition, this work studies some heuristics for sorting the list of virtual machines 
 (e.g., sorting by the earliest starting time, or latest finishing time, or the longest duration time first, etc.)
 to allocate VM.
 We evaluate the EM using CloudSim toolkit and jobs log-traces in the Feitelson's Parallel Workloads Archive.
Simulation's results show that all of EM-ST, EM-LFT and EM-LDTF 
algorithms could reduce  
  total energy consumption compared to state-of-the-art of power-aware VM allocation algorithms. 
 (e.g. Power-Aware Best-Fit Decreasing (PABFD) \cite{Beloglazov2011})).
 
\end{abstract}


%
%




\section{Introduction}
\label{sec:intro}
Cloud computing, which enables Infrastructure-as-a-Service (IaaS), provides users with computing
resources in terms of virtual machines (VMs)
to run their applications \cite{Garg2009,Le2011,Beloglazov2012,barroso2013datacenter}.
Infrastructure of cloud systems are built from virtualized data centers with thousands of high-performance computing servers
 \cite{Le2011,Beloglazov2011,Beloglazov2012,barroso2013datacenter,Garg2009,Vis2011}.
Power consumption in these data centers requires multiple Mega-watts \cite{Fan2007a,Le2011}. 
Le et al. \cite{Le2011} estimates the energy cost of a single data center is more than \$15M per year. 
The power consumption is increased 
with the increasing scale of these data centers.
Therefore, advanced scheduling techniques for
reducing energy consumption of these cloud systems are highly
concerned for any cloud providers to reduce energy cost. 
Increasing energy cost and the need of environmental sustainability addressing energy efficiency is a hot research topic in cloud systems.

Energy-aware placement of VMs in cloud computing is still challenging \cite{Garg2009,Le2011,Vis2011}.
Many previous works \cite{Beloglazov2011,Panigrahy2011} showed that the virtual machine allocation is NP-Hard.
There are several works that have been proposed to address the problem of 
energy-efficient scheduling of VMs in cloud data centers. 
Much research \cite{Beloglazov2012,Beloglazov2011,Panigrahy2011}
presents methodologies for 
consolidating virtual machines in cloud data centers by using 
bin-packing heuristics (such as First-Fit Decreasing \cite{Panigrahy2011}, and/or Best-Fit Decreasing \cite{Beloglazov2011}).
They attempt to minimize the number of running physical machines
and to turn off as many idle physical machines as possible. 
Consider a $d$-dimensional resource allocation
where each user requests a set of virtual machines (VMs). 
Each VM requires multiple resources (such as CPU, memory, and IO)
and a fixed quantity of each resource at a certain time interval.
Under this scenario, using a minimum of physical machines may not be a good solution.
Our observations show that
using a minimum number of physical machines
is not necessarily a good solution to minimize total energy consumption.
In a homogeneous environment where all physical servers are identical,
the power consumption of each physical server is linear to its CPU utilization,
i.e.,
a schedule with longer working time will consume more energy
than another schedule with shorter working time.

\begin{table}[!hbt]
\caption{This example shows that using a minimum number of physical servers is not optimal. (*: normalized demand resources to maximum host's total capacity resource)}
\label{table:example1}
\begin{center}
\scalebox{0.9}{
\begin{tabular}{|l|c|c|c|c|c|}
\hline
VM ID & CPU* & RAM* & Network* & Starttime & Duration \\
	&	&	&	&	& (hour) \\
\hline
VM1             & 0.5                             & 0.1                      & 0.2                            & 1                              & 20                                  \\
VM2             & 0.5                             & 0.3                      & 0.2                            & 1                              & 2                                  \\
VM3             & 0.2                             & 0.4                      & 0.2                            & 1                              & 1                                   \\
VM4             & 0.2                             & 0.4                      & 0.2                            & 1                              & 1                                   \\
VM5             & 0.1                             & 0.1                      & 0.3                            & 1                              & 1                                  \\
VM6             & 0.5                             & 0.3                      & 0.2                            & 2                              & 17                                  \\ \hline
\end{tabular}}
\end{center}
\end{table}

In our knowledge, our previous works \cite{HungGAPA,HungFDSE2014} are first works that consider energy-aware virtual machines allocation problem (EVMAP)
where a user requests VM with fixed starting time and non-preemptive during their duration time. 
In this paper, we study time-aware best-fit decreasing heuristics for the VM allocation.
We present here an example to demonstrate our ideas to 
minimize total energy consumption of all physical machines in the VM placement with fixed starting time and duration time.
For example, given six virtual machines (VMs) with their resource
demands described in Table~\ref{table:example1}. 
In the example, a bin-packing-based algorithm results in a schedule $S_{1}$
in which two physical servers are used:
one for allocating VM1, VM3, VM4, and VM5; and another one for
allocating VM2 and VM6. The total completion time of two servers
is (20 + 18) = 38 hours. However, in another schedule $S_{2}$ 
where VMs are placed on three physical servers,
VM1 and VM6 on the first physical server,
VM3, VM4 and VM5 on the second physical server,
and VM2 on the third physical server,
then the total completion time of the three servers is (20 + 1 + 2) = 23 hours.


In this paper, we propose an energy-aware, minimizing total busy time heuristic, denoted as EM.
The EM places VMs that request multiple resources in the fixed interval time 
 (e.g. fixed starting time and finishing time) and non-preemption in their during time
 into physical machines to minimize total energy consumption of all physical machines while meeting all resource requirements.
%
The EM allocates a new VM onto a physical machine that minimizes metric ranking, namely $RET$,
 which is included increasing total busy time and 
 remaining available resource on the mapping. 
 The EM uses resource utilization during executing time period of a physical machine
 as performance metrics 
 that is used to fully assign utilization each resource in a physical machine.
 We evaluate the EM in three methods of sorting list of VMs that denoted as EM-ST, EM-LFT, EM-LDTF.
Numerical simulations using CloudSim toolkit, we compare EM-ST, EM-LFT, EM-LDTF
 with Power-Aware Best-Fit Decreasing (PABFD) \cite{Beloglazov2011}, modified best-fit decreasing with sorting the list of VMs by their starting time (BFD-ST),
 and our previous algorithm (e.g. MinDFT-ST \cite{HungFDSE2014}).
 With two realistic log-traces in the Feitelson's Parallel Workloads Archive,
 our simulation results show that EM-ST, EM-LFT and EM-LDTF 
 could reduce average of 44\% and 45\% respectively
   total energy consumption compared with PABFD \cite{Beloglazov2011}) and BFD-ST.
   Moreover, EM-ST, EM-LFT and EM-LDTF have less total energy consumption than the MinDFT-ST, MinDFT-LFT and MinDFT-LDTF respectively.

The rest of this paper is structured as follows. 
Section \ref{sec:related} discusses related works.
Section \ref{sec:problem} describes the energy-aware VM allocation problem with multiple requested resources, fixed starting and duration time. 
We also formulate the objective of scheduling, and present our theorems. 
 The proposed EM algorithm is presented in Section \ref{sec:algoEMinRET}.
 Section \ref{sec:experiment} discusses our performance evaluation using simulations.
 Section \ref{sec:concl} concludes this paper and introduces future works.
\section{Related Works}
\label{sec:related}

Some other works proposed algorithms that consolidate VMs 
 onto a small set of physical machines (PMs) in virtualized datacenters \cite{Beloglazov2011,Beloglazov2012,Knauth2012,Chen2014} 
 to minimize energy/power consumption of PMs.
 Many works have considered the VM placement problem as a bin-packing problem,
and have used bin-packing heuristics to
place VMs onto a minimum number of PMs 
to minimize energy consumption \cite{Beloglazov2011,Beloglazov2012}. 
Microsoft research group \cite{Panigrahy2011} has studied first-fit decreasing (FFD) based heuristics for vector bin-packing to minimize number of physical servers in the VM allocation problem.
 Some other works also proposed meta-heuristic algorithms to minimize the number of physical machines.
 A hill-climbing based allocation of each independent VM is studied in
 \cite{Goiri2010}. 
 In the VM allocation problem, however, minimizing the number of used physical machines does not equal to minimizing the total energy consumption of all physical machines.

Beloglazov et al. 
 \cite{Beloglazov2011,Beloglazov2012} have proposed VM allocation problem as bin-packing problem and presented a power-aware best-fit decreasing
(denoted as PABFD) heuristic.
PABFD sorts all VMs in a decreasing order of CPU utilization and tends to
allocate a VM to an active physical server that would take the minimum
increase of power consumption. 
Knauth et al. \cite{Knauth2012} proposed the OptSched scheduling algorithm to 
 reduce cumulative machine up-time (CMU) by 60.1\% and 16.7\% in comparison to a round-robin and First-fit.
 The OptSched uses an minimum of active servers to process a given workload.
 In a heterogeneous physical machines, the OptSched maps a VM to a first available and the most powerful machine that has enough VM's requested resources. Otherwise, the VM is allocated to a new unused machine. 
These previous works do not consider multiple resources, fixed starting time and non-preemptive duration time of these VMs.
Therefore,
it is unsuitable for the power-aware VM allocation
considered in this paper,
i.g. these previous solutions can not result in a minimized total energy consumption
for VM placement problem with certain interval time while still
fulfilling the quality-of-service.

Some other research \cite{Garg2009,Le2011} considers
HPC applications (or HPC jobs) in HPC clouds. 
Garg et al. \cite{Garg2009} proposed a meta-scheduling problem
to distribute HPC applications to cloud systems with distributed
$N$ data centers. The objective of scheduling is minimizing $CO_{2}$
emission and maximizing the revenue of cloud providers.
Le et al. \cite{Le2011} distribute VMs across distributed
cloud virtualized data centers whose electricity prices
are different in order to reduce
the total electricity cost.
Our proposed EM algorithm that
differs from these previous works. Our EM algorithm use the VM's fixed
starting time and duration time to minimize the total working
time on physical servers, and consequently minimize the total
energy consumption in all physical servers. 
In my knowledge, our solution is original from any previous works include surveyed works in \cite{BelBuLZA2010Taxonomy,orgerie2014survey,Hameed2014}.


In 2007, Kovalyov et al. \cite{kovalyov2007fixed} has presented a work describing characteristics of a fixed interval scheduling problem in that each job has fixed starting time, fixed processing time, and is only processed in the fixed duration time on a available machine. The scheduling problem can be applied in other domains. 
Angelelli et al. \cite{Angelelli20113650} considered interval scheduling with a resource constraint in parallel identical machines. The authors presented that if the number of constraint resources in each parallel machine is fixed value $R \geqslant 2$, the decision problem is strongly NP-Complete.
\section{Problem Description}
\label{sec:problem}
\subsection{Notations}
We use the following notations in this paper:

$ vm_{i}$: The $i^{th}$ virtual machine to be scheduled.

$ M_{j}$: The $j^{th}$ physical server.

$ S $: A feasible schedule. 

$ P_{j}^{idle}$: Idle power consumption of the $M_{j}$.

$ P_{j}^{max} $: Maximum power consumption of the $M_{j}$.

$ P_{j}(t) $: Power consumption of a single physical server ($M_{j}$) at a time point $t$.


$ ts_{i}$: Fixed starting time of $vm_{i}$.

$ dur_{i}$: Duration time of $vm_{i}$.

$ T $: The maximum time of the scheduling problem.

$ n_{j}(t) $: Set of indexes of all VMs that are assigned to the physical machine $M_{j}$ at time $t$.

$T_{j}$ : The working time of a physical server.

 $ e_{i} $: The energy consumption for running the $ vm_{i}$ in the physical machine where the $ vm_{i}$ is allocated.\\

\subsection{Power consumption model}
In this paper, we use the following energy consumption model proposed in \cite{Fan2007a} for a physical machine.
 The power consumption of the $M_{j}$, denoted as $ P_{j}(.)$, is formulated as follow:
\begin{equation}
\label{eq:power}
 P_{j}(t) = P_{j}^{idle} + (P_{j}^{max} - P_{j}^{idle}) U_{j}(t) 
 \end{equation}
 
 
 The CPU utilization of the physical server at time $t$, denoted as $ U_{j}(t)$, is defined as the average percentage of total allocated 
 computing powers of $n_j(t)$ VMs that is allocated to the $M_j$. 
We assume that all cores in CPU are homogeneous, i.e. $\forall c=1,2,...,PE_j: MIPS_{j,c}=MIPS_{j,1}$ , The CPU utilization is formulated as follow:
 \begin{equation}
 \label{eq:cpuutilization}
 U_{j}(t) = (\dfrac{1}{PE_{j} \times MIPS_{j,1}}) \sum_{c=1}^{PE_{j}} \sum_{i  \in  n_j(t)} mips_{i,c} 
 \end{equation}

The energy consumption of the server in the period of time [$t_1, t_2$] is formulated as follow:
\begin{equation}
\label{eq:energy}
E_{j} = \int_{t_{1}}^{t_{2}} P_{j}( U_{j}(t)) dt
\end{equation}

{\noindent}where: \\
 $ U_{j}(t) $ : CPU utilization of $M_{j}$ at time $ t $ and $ 0 \leq U_{j}(t) \leq 1 $. \\
 $ PE_{j}$ : Number of processing elements (i.e. cores) of the $M_{j}$.\\
 $ mips_{i,c} $	: Allocated MIPS of the c$^{th}$ processing element to the $vm_{i}$ by the $M_{j}$.\\
 $ MIPS_{j,c} $ : Maximum capacity computing power (Unit: MIPS) of the c$^{th}$ processing element on the $M_{j}$.\\


\subsection{Energy-Aware VM Allocation Problem with Fixed Interval and Non-preemption}

We define the energy-aware VM allocation problem with fixed starting time and non-preemption (EVMAP) as following:

Given a set of virtual machines $ vm_{i}$ ($i = 1,2,...,n$) to be scheduled on a
set of physical servers $ M_{j}$ ($j = 1,2,...,m$). 
Each VM is represented
as a d-dimensional vector of demand resources, 
i.e. $vm_{i} = (x_{i,1}, x_{i,2}, ..., x_{i,d}) $.
Similarly,
each physical machine is denoted as a d-dimensional 
vector of capacity resources, 
i.e. $ M_{j} = (y_{j,1}, y_{j,2}, ...,  y_{j,d}) $.
We consider types of resources such as processing element (PE), 
 computing power (Million instruction per seconds -MIPS),
 physical memory (RAM), network bandwidth (BW), and storage. 
Each $vm_{i}$ is started at a fixed starting time ($ts_{i}$) 
and is non-preemptive during its duration time ($dur_{i}$). 

The energy consumption of a physical machine in a unit of time is denoted as $E_j^{0}$.
The energy consumption of each VM is denoted as $e_i$.
The objective is to find out a feasible schedule $S$ that minimizes
the total energy consumption
in the equation (\ref{eq:minimize}) 
 with $i \in \{1,2,...,n\}$, $j \in \{1,2,...,m\}$, $t \in [0;T]$ as following:


\begin{equation}
\label{eq:minimize}
\textbf{Minimize} \   ( E_j^{0} \times \sum_{j=1}^{m} T_{j}  +  \sum_{i=1}^{n} e_{i} )
\end{equation}  


{\noindent}where
the working time of a physical server, denoted as $T_{j}$, is defined as union of interval times of all VMs 
 that are allocated to a physical machine $j^{th}$ at time $t$. 

%
%




The scheduling problem has the following hard constraints that are described in our previous work \cite{HungFDSE2014}.
%
%
%
%
%
%

\section{EM: Energy-Aware, Minimizing Total Busy Time Heuristic}
\label{sec:algoEMinRET}
\subsection{Scheduling algorithm}
\label{sec:schedalgo}

\begin{algorithm*}[!ht]
\caption{EM: Energy-Aware, Minimizing Total Busy Time Heuristic}\label{alg:eminret}
 \begin{algorithmic}[1]
\Function{EM}{}
\State	\textit{Input:} vmList - a list of virtual machines to be scheduled
\State	\textit{Input:} hostList - a list of physical servers
\State	\textit{Output:} mapping (a feasible schedule) or null 
\State	vmList = sortVmListByOrder( vmList, order=[starttime, finishtime] ) \Comment{1}\label{line:sortvmlistbycriteria}

\For {$j=1$ to m} \State T[j] = 0 \EndFor
\For {$i = 1$ to n} \Comment{Loop on the VMs list}
		\State 	vm = vmList.get(i) 
		\State	allocatedHost = null 
		\State	T1 = sumHostsTotalBusyTime( T ) \Comment{Calculating the total busy time of all active physical servers:}
		\State	minRETime = $+\infty$
		\For {$j = 1$ to m}  \Comment{Loop on the hosts list}
         	\State host = hostList.get(j)
         	\State hostVMList = sortVmListByOrder( host.getVms(), order=[starttime, finishtime]) 
			\If {isUsed(host) and host.checkAvailableResource( vm )} \\
				\Comment{ host's available resources has enough the vm's requested resources } 				
				\State preTime = T[ host.id ] 
				\State	host.vmCreate(vm) \Comment{begin test }
				\State T[ host.id ] = estimateHostTotalBusyTime(host);
				\State T2 = sumHostsTotalBusyTime( T[] )
				\State host.vmDestroy(vm) \Comment{end test}
				\State RETime = EstimateMetricTimeResEff(  T2 - T1, host ) \Comment{the function is invoked to estimate the metric for increasing time and resource efficiency}
				\If {(minRETime $>$ RETime )}
					\State bestEMinRET = t
					\State	minRETime = RETime 
					\State	allocatedHost = host	
				\EndIf
				\State T[ host.id ] = preTime
				\Comment{ Next iterate over hostList and choose the host that minimize the value of different time and resource efficiency } 
         \EndIf
		\EndFor \Comment{ End for host list } 
		
		\If {(allocatedHost = null) }
			\State allocatedHost = \{Select a new host that has maximum of TotalMIPS/Watts from idle hosts in the $hostList$ \}
		\EndIf
		\If {(allocatedHost != null) }
			\State allocate the $vm$ to the $host$
			\State add the pair of $vm$ (key) and $host$ to the $mapping$
		\EndIf
\EndFor	\Comment{end for vm list }
\State	return $mapping$ 
\EndFunction

\State {sumHostsTotalBusyTime}({T[]}) = $\sum_{j=1}^{m} T_{j}$ \Comment{T[1...m]: Array of total completion times of $m$ physical servers}

\end{algorithmic}
\end{algorithm*}

\begin{algorithm*}[ht]
\caption{Estimating the metric for increasing time and resource efficiency}\label{alg:estimretime}
 \begin{algorithmic}[1]
 \Function{EstimateMetricTimeResEff}{}
 \State	\textit{Input:} $t^{diff}$ - a different time before and after allocation the VM.
 \State	\textit{Input:} host - a candidate physical machine.
 \State	\textit{Output:} ret - a value of metric time and resource efficiency 
 \State Set $\mathcal{R}$=\{cpu, ram, netbw, io, storage, time\}
 \State $j$=host.getId(); $n_j$=host.getVMList();
 \For{$r \in \mathcal{R}$} 
 	\State Calculate the resource utilization, $U_{j,r}$ as in the Equaltion (\ref{eq:resutilization}). 
 \EndFor	
 \State $weights[] \leftarrow$ Read resource weights from configuration file.
 \If {($t^{diff} \neq 0$)}
 	\State $ret = t^{diff} \times weights[time]  \sqrt{\sum_{r \in \mathcal{R}}  (weights[r] \times (1-U_{j,r}))^2}$  \Comment{weights[time] is weight of the different time}
 	\Else
		\State $ret = \sqrt{\sum_{r \in \mathcal{R}}^{} ((1 - U_{j,r}) \times weights[r])^2} $ 
  \EndIf
 \State return $ret$
 \EndFunction 
 
\end{algorithmic}
\end{algorithm*}

In this section, we present our energy-aware scheduling algorithm named as EM.
The EM algorithm presents a metric to unify the increasing time and estimated resource efficiency when mapping a VM onto a physical machine.
Then, EM will choose a host that has the minimum of the metric. 
   Our previous MinDFT-ST/FT \cite{HungFDSE2014} only focused on minimizing the increasing time when mapping a VM onto a physical machine.
   The EM additionally considers resource efficiency
   during an execution period of a physical machine in order to fully utilize resources in a physical machine.

Based the Equation \ref{eq:cpuutilization}, the utilization of a resource $r$ (resource $r$ can be CPU, physical memory, network bandwidth, storage, etc.) of a PM $j$-th, denoted as $U_{j,r}$, is formulated as:
 \begin{equation}
 \label{eq:resutilization}
 U_{j,r} = \sum\limits_{s \in n_{j}} \dfrac{V_{s,r}}{ H_{j,r}}.
  \end{equation}

{\noindent}where $n_{j}$ is the list of VMs that are assigned to the physical machine $j$, 
 $V_{s,r}$ is the amount of requested resource $r$ of the VM $s$ (note that in our problem the $V_{s,r}$ is the fixed value in each user request), 
 and $H_{j,r}$ is the maximum capacity of resource $r$ in the physical machine $j$.
  
Inspired by research from Microsoft research team \cite{Panigrahy2011,Chen2014},
 resource efficiency of a physical machine $j$-th, denoted by $RE_{j}$, 
 is Norm-based distant \cite{Panigrahy2011} of two vectors: normalized resource utilization vector and unit vector $\mathbf{1}$, 
 the resource efficiency is formulated as:
\begin{equation}
\label{eq:resefficiency}
RE_{j} = \sqrt{ \sum_{r \in \mathcal{R}} ((1 - U_{j,r}) \times w_r)^2 }
 \end{equation}
{ \noindent}where $\mathcal{R}$=\{MIPS, memory, netbw, storage\} is the set of resource types in a host, $w_r$ is the weight of resource $r$ in a physical machine.

 In this paper, we propose a unified metric for increasing time and resource efficiency that is calculated as:
 \begin{equation}
 \label{eq:retime}
 RET = \begin{cases}
 		t^{diff} \times w_{r=time} \times RE_j, & \text{if $t^{diff} \neq 0$}.\\
		RE_j, & \text{otherwise}.
 	\end{cases} 
 \end{equation}
  
  The EM chooses an used host that has the minimum value of the $RET$ metric to allocate for a new VM.
  If the EM could not find out any used host to allocate the new VM, then the EM will open a new host that has maximum 
  ratio of performance-per-watts, which is calculated as total MIPS in all cores and maximum power consumption.


We present pseudo-code for our proposed EM in Algorithm~\ref{alg:eminret}.
 The EM can sort the list of VMs by earliest starting time first, or earliest finishing time first, or longest duration time first, etc..
The EM solves the scheduling problem in time complexity of $\BigO{n \times m \times q}$
where $n$ is the number of VMs to be scheduled, $m$ is the number of physical machines, and $q$ is the maximum number of allocated VMs in the physical machines $M_j, \forall j=1,2,...,m$.
\section{Performance Evaluation}
\label{sec:experiment}
\subsection{Algorithms}

In this section, we study the following VM allocation algorithms:

\begin{itemize}
\item
PABFD, a power-aware and modified best-fit decreasing heuristic \cite{Beloglazov2011}\cite{Beloglazov2012}.
The PABFD sorts the list of $VM_{i}$ (i=1, 2,..., n) by
their total requested CPU utilization, and 
assigns a new VM to any host that has a minimum increase in power consumption.

\item
BFD-ST, a modified best-fit decreasing heuristic. The BFD-ST sorts the list of VMs by their starting times and 
assigns a new VM to any host that has a minimum increase in power consumption (i.e., similar to the PABFD).



\item
MinDFT-ST, MinDFT-LFT and MinDFT-LDTF: 
 The MinDFT-ST \cite{HungFDSE2014} sorts the list of VMs by VM's earliest starting time first and
 allocates each VM (in a given set of VMs) to a host that has a minimum increase in the total completion time of hosts.
 Both of MinDFT-LFT and MinDFT-LDTF use core MinDFT algorithm \cite{HungFDSE2014}. 
 Instead of sorting list of virtual machines by their earliest starting time first as in MinDFT-ST, MinDFT-LFT and MinDFT-LDTF 
 sort the list of VMs by their latest finished time and longest duration time first respectively.

\item
EM (denoted as EM in the chart of energy consumption) is our proposed algorithm discussed in Section \ref{sec:schedalgo}. We evaluate the EM with three  configurations:
The EM-ST sorts the list of VMs by VM's earliest starting time first and host's allocated VMs by its finishing time.
The EM-LFT sorts the list of VMs by VM's latest finishing time first and host's allocated VMs by its finishing time.
The EM-LDTF sorts the list of VMs by VM's longest duration time first and host's allocated VMs by its finishing  time.

\end{itemize}

\subsection{Methodology}

\begin{table}[htp]
\caption{Eight (08) VM types in simulations}
\label{tab:vmtype}


\centering
\scalebox{0.9}{
\begin{tabular}{|l|r|r|r|r|r|}
\hline
VM Type                    & \multicolumn{1}{l|}{MIPS} & \multicolumn{1}{l|}{Cores} & \multicolumn{1}{l|}{Memory} & \multicolumn{1}{l|}{Net. Bw.} & \multicolumn{1}{l|}{Storage} \\ 
                    &  &  & (MBytes) & (Mbits/s) & (GBytes) \\ 
\hline
Type 1 & 2500 & 8 & 6800 & 100 & 1000\\
Type 2 & 2500 & 2 & 1700 & 100 & 422.5\\
Type 3 & 3250 & 8 & 68400 & 100 & 1000\\
Type 4 & 3250 & 4 & 34200 & 100 & 845\\
Type 5 & 3250 & 2 & 17100 & 100 & 422.5\\
Type 6 & 2000 & 4 & 15000 & 100 & 1690\\
Type 7 & 2000 & 2 & 7500 & 100 & 845\\
Type 8 & 1000 & 1 & 1875 & 100 & 211.25\\ \hline
\end{tabular}}
\end{table}

\begin{table}[htp]
\caption{Information of three typical physical machines (Hosts)} 
\label{tab:Hostinfo}
\centering
\scalebox{0.9}{
\begin{tabular}{|c|r|r|r|r|r|}
\hline
Type	& \multicolumn{1}{l|}{MIPS} & \multicolumn{1}{l|}{Cores} & \multicolumn{1}{l|}{Memory} & \multicolumn{1}{l|}{Net. Bw.} & \multicolumn{1}{l|}{Storage} \\ 
                    &  &  & (MBytes) & (Mbits/s) & (GBytes) \\  \hline
M1 & 3250 & 4 & 30720 & 10000 & 10000  \\ 
M2 & 3250 & 16 & 140084 & 10000 & 10000 \\ 
M3 & 2500 & 16 & 14336 & 10000 & 10000   \\ \hline
\end{tabular}}
\end{table}

\begin{table}[htp]
\caption{Host power consumption of three typical servers} 
\label{tab:HostPower}
\centering
\scalebox{0.9}{
\begin{tabular}{|c|c|c|}
\hline
\multicolumn{1}{|c}{Host Type}& \multicolumn{1}{|c|}{$P^{idle}$} & \multicolumn{1}{c|}{$P^{max}$} \\ \hline
M1 & 210 & 300 \\ 
M2 & 420 & 600 \\ 
M3 & 350 & 500 \\ \hline
\end{tabular}}
\end{table}

\begin{table*}[htp]
\centering
\caption{Result of simulations using the first 400 HPC jobs that included 7495 VMs of the HPC2N Seth log-trace \cite{HPC2NWorkload}.} 
\label{table:simresult-hpc2n}
\begin{tabular}{|l|c|c|c|c|c|}
\hline
Algorithm & \multicolumn{1}{l|}{\#Hosts} & \multicolumn{1}{l|}{\#VMs} & \multicolumn{1}{l|}{Energy (KWh)} & \multicolumn{1}{l|}{Normalized Energy} & \multicolumn{1}{l|}{Energy Saving (\%)} \\ 
\hline
PABFD \cite{Beloglazov2011} (baseline)  & 5000 & 7495 & 6331.58 & 1.00 & 0\% \\
BFD-ST & 5000 & 7495 & 6162.47 & 0.97 & 3\% \\
MinDFT-ST & 5000 & 7495 & 6106.39 & 0.96 & 4\% \\
MinDFT-LFT & 5000 & 7495 & 4163.23 & 0.66 & 34\% \\
MinDFT-LFT & 5000 & 7495 & 5551.55 & 0.88 & 12\% \\
EM-ST & 5000 & 7495 & 3649.28 & 0.58 & 42\% \\
EM-LFT & 5000 & 7495 & 3368.62 & 0.53 & 47\% \\
EM-LDTF & 5000 & 7495 & 3119.19 & 0.49 & 51\%\\
\hline
\end{tabular}
\end{table*}

\begin{table*}[htp]
\centering
\caption{Result of simulations using the first 100 HPC jobs that included 12536 VMs of SDSC BLUE log-trace \cite{SDSCBLUEWorkload}.} 
\label{table:simresult-sdscblue}
\begin{tabular}{|l|c|c|c|c|c|}
\hline
Algorithm & \multicolumn{1}{l|}{\#Hosts} & \multicolumn{1}{l|}{\#VMs} & \multicolumn{1}{l|}{Energy (KWh)} & \multicolumn{1}{l|}{Normalized Energy} & \multicolumn{1}{l|}{Energy Saving (\%)} \\ \hline
PABFD  \cite{Beloglazov2011} (baseline) & 10000 & 12536 & 4173.36 & 1.00 & 0\% \\
BFD-ST & 10000 & 12536 & 4416.85 & 1.06 & -6\% \\
MinDFT-ST & 10000 & 12536 & 4397.02 & 1.05 & -5\% \\
MinDFT-LFT & 10000 & 12536 & 3351.88 & 0.80 & 20\% \\
MinDFT-LDTF & 10000 & 12536 & 4046.22 & 0.97 & 3\% \\
EM-ST & 10000 & 12536 & 2434.41 & 0.58 & 42\% \\
EM-LFT & 10000 & 12536 & 2437.64 & 0.58 & 42\% \\
EM-LDTF & 10000 & 12536 & 2437.64 & 0.58 & 42\%\\
\hline
\end{tabular}
\end{table*}

\begin{figure}[!htb]
    \centering
	\includegraphics[width=0.45\textwidth,height=4.5cm]{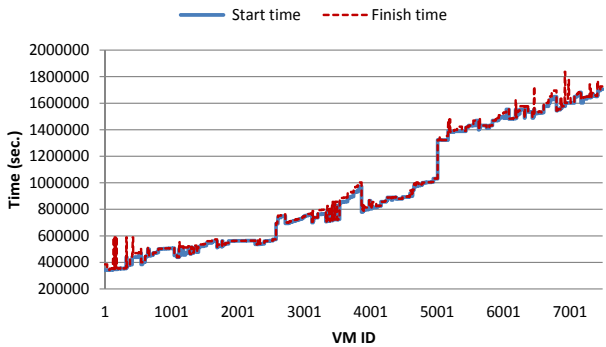}
    \caption[VM's starting and finishing time]{Starting time (blue line) and finishing time (red dotted line) of VMs in simulations with HPC2N Seth log-trace \cite{HPC2NWorkload}.}
    \label{fig:vmtime}
\end{figure}

\begin{figure}[!htb]
\centering
\includegraphics[width=0.45\textwidth,height=4.8cm]{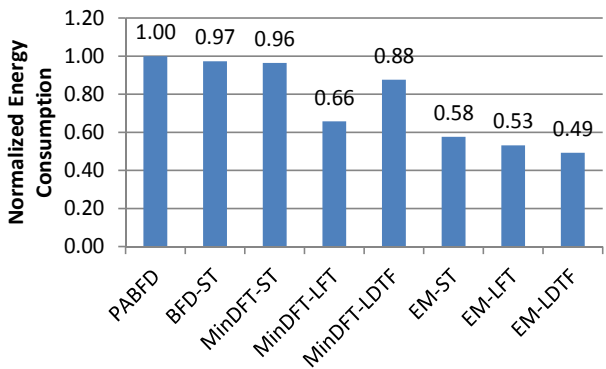}
\caption{Result of simulations with HPC2N Seth log-trace \cite{HPC2NWorkload}.}
    \label{fig:scaledenergyresult-hpc2n}
\end{figure}

\begin{figure}[!htb]
    \centering
    \includegraphics[width=0.45\textwidth,height=4.5cm]{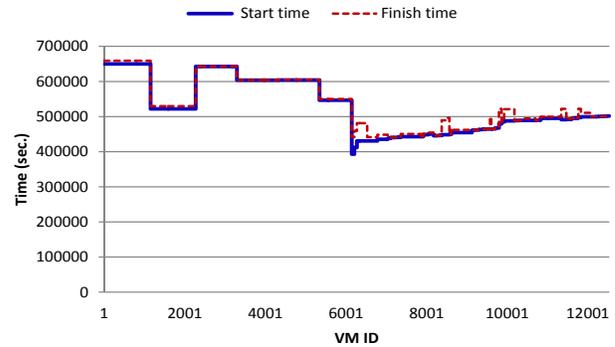}
    \caption[VM's starting and finishing time]{Starting time (blue line) and finishing time (red dotted line) of VMs in simulations with SDSC Blue Horizon log-trace \cite{SDSCBLUEWorkload}.}
    \label{fig:vmtime-sdscblue}
\end{figure}

\begin{figure}[!htb]
\centering
\includegraphics[width=0.45\textwidth,height=4.5cm]{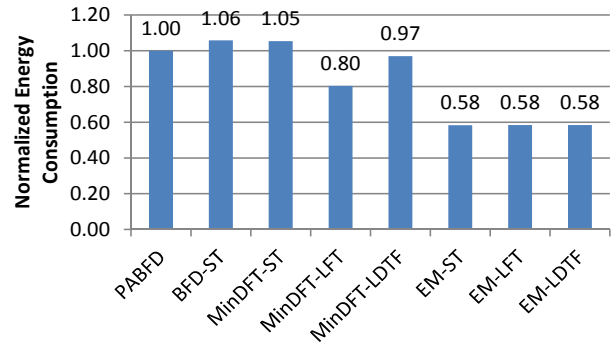}
\caption{Energy consumption (Unit: KWh). Result of simulations with SDSC Blue Horizon log-trace \cite{SDSCBLUEWorkload}.}
    \label{fig:chart-scaledenergy-sdsc-blue}
\end{figure}

We evaluate these algorithms by simulations using the CloudSim \cite{Cloudsim} to create a simulated
 cloud data center system that has some thousands of physical machines, heterogeneous VMs,
 and with thousands of CloudSim's cloudlets \cite{Cloudsim} (we assume that each
HPC job's task is modeled as a cloudlet that is run on a single VM). 
The information of VMs (and also cloudlets) in these simulated workloads 
is extracted from a real log-trace ( 
HPC2N Seth log-trace \cite{HPC2NWorkload}) 
in Feitelson's Parallel Workloads Archive (PWA) \cite{PWA} to 
model HPC jobs. When being converted from the 
log-trace, each cloudlet's length is a product of the system's 
processing time and CPU rating (we set the CPU rating is equal to included VM's MIPS).
 We convert job's submission time, job's start time (if the start time is missing, then the start time is equal to sum of job's submission time and job's waiting time), 
 job's request run-time, 
 and job's number of processors 
 in job data from the log-trace in the PWA
 to VM's submission time, starting time and duration time, 
 and number of VMs (each VM is created in round-robin in the four types of VMs in Table \ref{tab:vmtype} on the number of VMs). 
 Eight (08) types of VMs as presented in the Table \ref{tab:vmtype} are similar to categories in Amazon EC2's 
 VM instances such as high-CPU and low-memory VM (Type 1), high-CPU and high-memory VM (Type 3-4), high-memory VM (Type 5-6), normal VM (Type 2, Type 7), and micro VM (Type 8). 
 All physical machines are divided into three group 
 and each physical machine has resources as one of host type in the Table \ref{tab:Hostinfo} and 
 its power model is shown in Table \ref{tab:HostPower}.
In the simulations, we use weights as following:
(i) weight of increasing time of mapping a VM to a host: \{0.001, 0.01, 1, 100, 3600\};
(ii) weights of computing resources such as number of MIPS per CPU core, physical memory (RAM), network bandwidth, and storage respectively: 940, 24414, 1, 0.0001 respectively. 
We will discuss how to choose these values for weights of resources in another paper. 
We simulate on combination of these weights. 
Each EM's total energy consumption is average of five times simulation with various weights of increasing time (e.g. 0.001, 0.01, 1, 100, or 3600).

We choose PABFD \cite{Beloglazov2011} as the baseline algorithm 
because the PABFD is a famous power-aware best-fit decreasing
in the energy-aware scheduling research community.
We also compare our proposed VM allocation algorithm with modified best-fit decreasing (BFD) to show 
the importance of with/without considering VM's starting time and finish time 
in reducing the total energy consumption of VM placement problem.


\subsection{Results and Discussions}
\label{sec:resultsanddiscussions}

The simulation results are shown in the two following tables and figures.
  The two figures, Fig. \ref{fig:vmtime} and Fig. \ref{fig:vmtime-sdscblue}, show charts of starting times and finishing times
  of the VMs in a simulation (the simulations have the same starting times and duration times of VMs).
Table \ref{table:simresult-hpc2n} shows simulation results of scheduling algorithms solving scheduling problems 
 with 7495 VMs and 5000 physical machines (hosts),
 in which VM's data is converted from the HPC2N Seth log-trace \cite{HPC2NWorkload}.
 Table \ref{table:simresult-sdscblue} shows simulation results of scheduling algorithms solving scheduling problems 
  with 12536 VMs and 10000 physical machines (hosts),
  in which VM's data is converted from the SDSC BLUE log-trace \cite{SDSCBLUEWorkload}.
Both Figure \ref{fig:scaledenergyresult-hpc2n} and Figure \ref{fig:chart-scaledenergy-sdsc-blue} show bar charts comparing energy consumption of VM allocation algorithms that scale with the PABFD.
None of the algorithms use VM migration techniques,
and all of them  satisfy the Quality of Service
(e.g. the scheduling algorithm provisions the maximum of user VM's requested resources).
We use total energy consumption as the performance metric for
evaluating these VM allocation algorithms.
The energy saving shown in both Table \ref{table:simresult-hpc2n} and Table \ref{table:simresult-sdscblue} is the reduction of
total energy consumption of the corresponding algorithm
compared with the baseline PABFD  \cite{Beloglazov2011} algorithm.

Table \ref{table:simresult-hpc2n} shows that, compared with PABFD \cite{Beloglazov2011} and BFD-ST,
 our EM with three configurations (denoted as EM-ST, EM-LFT, EM-LDTF) 
 can reduce the total energy consumption by average 47\% and 45\% respectively in simulations with the first 400 jobs of the HPC2N Seth log-trace. 
 Table \ref{table:simresult-sdscblue} shows that, compared with PABFD \cite{Beloglazov2011} and BFD-ST,
 all of our EM-ST, EM-LFT and EM-LDTF
 reduce total energy consumption by average 42\% and 45\% respectively
 in simulations with the first 100 jobs in the SDSC BLUE log-trace. 
In summary, three configurations of the EM algorithm 
 allocate all VMs onto physical machines using less total energy consumption than both PABFD and BFD-ST.
 The EM has also less total energy consumption than MinDFT (with same method of sorting of VMs list).
 Moreover, the EM-LDTF has the minimum of total energy consumption in the results of simulations.
\section{Conclusion and Future Work}
\label{sec:concl}
In this paper, we formulated an energy-aware VM allocation problem with fixed interval time and non-preemption, denoted as EVMAP.
We also discussed our key observation in the VM allocation problem.
 Using bin-packing heuristics (such as PABFD, BFD-ST) to reach the minimum of used physical machines (PMs) 
 could not minimize total energy consumption of PMs in the EVMAP.
 Minimizing the sum of total busy time of all PMs could imply minimizing total energy consumption of PMs in the EVMAP.
Based on these observations,
we proposed EM algorithm to solve the EVMAP.


Our proposed EM and its sorting list of VMs by starting time (or longest duration time first, or latest finishing time first)
can all reduce the total energy consumption
of the physical servers 
compared with other algorithms in 
simulation results on two real HPC log-trace, which are the HPC2N Seth \cite{HPC2NWorkload} 
and SDSC Blue Horizon log-traces \cite{SDSCBLUEWorkload}, in the Feitelson's Parallel Workloads Archive \cite{PWA}. 
The combination of EM with its sorting list of VMs by longest duration time first (EM-LDTF) has the minimum of total energy consumption
 in the results of simulations. 

In future, we are developing EM into a cloud resource management software (e.g. OpenStack Nova Scheduler).
We are studying how to choose the right weights of time and resources 
(e.g. computing power, physical memory, network bandwidth, etc.) in Machine Learning techniques.


\section*{Acknowledgment}
This research is also funded by Vietnam National University Ho Chi Minh - University of Technology (HCMUT) under grant number T-KHMT-2015-33.

\balance
\bibliography{refs}  



\end{document}